\documentclass[fleqn,usenatbib]{mnras}

\usepackage{newtxtext,newtxmath,comment}
\usepackage{bm}

\usepackage[T1]{fontenc}
\usepackage{ae,aecompl}

\usepackage{graphicx}	
\usepackage{amsmath}	

\usepackage{amssymb}	
\usepackage[normalem]{ulem}
\usepackage{graphicx}
\usepackage{subcaption}

\title{A Response to paper Critical Evaluation of Studies Alleging Evidence for Technosignatures in the POSS1-E Photographic Plates by Watters et al. (2026)}

\author[Villarroel et al.]{
\and
Beatriz Villarroel,$^{1}$\thanks{E-mail: beatriz.villarroel@su.se}, Alina Streblyanska,$^{2}$
Stephen Bruehl,$^{3}$, Stefan Geier$^{4}$
\\
$^{1}$Nordita, KTH Royal Institute of Technology and Stockholm University, Hannes Alfvéns väg 12, SE-106 91 Stockholm, Sweden\\
$^{2}$Society of UAP Studies\\
$^{3}$ Department of Anesthesiology, Vanderbilt University Medical Center, 701 Medical Arts Building, 1211 Twenty-First Avenue South, Nashville, TN 37212 USA\\
$^{4}${Gran Telescopio Canarias (GRANTECAN), Cuesta de San Jos\'{e} s/n, 38712 Bre\~{n}a Baja, La Palma, Spain}
}

\date{Accepted XXX. Received YYY; in original form ZZZ}

\pubyear{2025}

\begin{document}
\label{firstpage}
\pagerange{\pageref{firstpage}--\pageref{lastpage}}
\maketitle

\begin{abstract}
We respond to the critique by \cite{Watters2026} of the statistical analyses in \cite{VillarroelPASP} and \cite{Bruehl2025}. We argue that the critique conflates object-level validation with ensemble-level statistical inference and relies on a reduced, heterogeneously filtered subset originally constructed for a different scientific purpose. This is evidenced by the strongly spatially skewed dataset used by \cite{Watters2026}, which differs significantly from the datasets used in recent VASCO studies. The pronounced heterogeneity and bias in sky coverage alone in \cite{Watters2026} produces an inherent excess of transients in the Earth’s shadow. We further question whether the aggressively filtered subset used in \cite{Watters2026} demonstrates a meaningful improvement in sample purity, given the twenty-fold reduction in sample size. A simple, visual check does not suggest that it does. An elaborate comparision of the fraction of visual transients in the two samples using a machine learning classifier, indeed suggests the aggressively filtered sample has a larger fraction of false detections. The aggressively filtered sample subset further lacks complete temporal information and is seriously statistically underpowered for testing the reported Earth-shadow deficit. We emphasise that the horizontal separation calculation used for plate assignment and time reconstruction as in \cite{Watters2026} depends on the inclusion of the cos(Dec) factor to ensure geometric consistency. Any omission would alter plate assignment and inferred observation times. Moreover, the analyses presented in \cite{Watters2026} do not include uncertainty estimates or error propagation, limiting the interpretability of the claimed null results. We conclude that the principal findings reported in \cite{VillarroelPASP} and \cite{Bruehl2025} are not invalidated by the analyses presented in \cite{Watters2026}.

\end{abstract}




\section{Commentary}
The Vanishing \& Appearing Sources during a Century of Observations (VASCO) project \citep{Villarroel2020} was originally designed to search for genuinely vanishing stars—objects that disappear without leaving any trace behind. Despite tracing 600 million sources, no such real vanishing star was identified by the project \citep{Solano2022}. 

Instead, the analysis uncovered thousands of short-lived flashes in historical photographic plates from the Palomar Observatory, exposed prior to the launch of Sputnik~I \citep{Villarroel2021, Solano2022}. Many of these events were initially examined as candidates for vanishing stars \citep{Villarroel2022b}, but were subsequently found to be brief flashes appearing in single exposures. A subset of these faint flashes appears in spatial groupings, raising the question of whether they represent genuine astronomical sources or some form of locally produced contamination \citep{Villarroel2021}. At the same time, certain cases, such as Solano’s Triple Transient \citep{Solano2024}, constitute compelling examples of high-confidence groupings. Motivated by this ambiguity, we have been developing methods to distinguish between the simple case of random, star-like defects and alignments of these objects, indicative of technosignature reflections \citep{Villarroel2022a}.

The findings of the past year have been particularly striking. We have identified a pronounced deficit of these transients within the Earth’s umbra at an altitude of approximately 42{,}000~km, significant at the $\sim 7.6\sigma$ level when the survey sky coverage is taken into account \citep{VillarroelPASP}. The observed deficit provides strong support for a solar-reflection origin of a meaningful subset of observed transients. In addition, we have found correlations at roughly the $\sim 3\sigma$ level between the occurrence of these events and periods associated with anomalous aerial phenomena and nuclear weapon tests \citep{Bruehl2025}.

Together, these results challenge a long-standing and deeply rooted assumption that most unrepeatable point sources in photographic plates can be attributed to dust specks, emulsion flaws, or scanning artefacts—although a substantial fraction undoubtedly can. Rather, the findings strongly suggest that, for at least a century, astronomers have overlooked a significant population of near-Earth objects concealed within the photographic plate material. An analogous population would today be difficult to disentangle from the multitude of optical glints produced by human-made space debris and satellites.

A recent article by \cite{Watters2026} uses samples from \cite{Solano2022} to argue that (i) the analysed datasets suffer from inconsistent definitions and insufficient validation, and that no feature has been demonstrated to constitute a verified optical transient; (ii) a substantial fraction of the analysed features cannot be reliably distinguished from catalogue stars, plate emulsion defects, or scan-related artefacts; (iii) the reported temporal correlations between feature detections and nuclear tests or UAP sighting reports become insignificant after appropriate normalisation and are dominated by the telescope observation schedule; and (iv) the reported deficit of features within the Earth’s shadow is not present when restricting the analysis to the most highly filtered datasets.

We do not address the claims of data inconsistencies raised in \cite{Watters2026}, as they rest on assumptions about sample construction that could have been clarified through a simple, collegial email to the first authors. Such an exchange would also have precluded the use of published datasets lacking time and date information for the transient detections and, most importantly, the use of a dataset constructed for an entirely different research purpose and where statistical completeness was not the aim.

We begin by examining the critique itself. Its central claim is that, because individual transients cannot be demonstrated to represent genuine astronomical observations, the analysis as a whole is therefore invalid. Watters’ critique implicitly assumes that object-level validation is a prerequisite for ensemble inference. That assumption is false in many fields of science, e.g. statistical mechanics, cosmology, particle physics, survey astronomy, and epidemiology. If we were to disqualify a statistical study due to false positives in the sample, all modern physics experiments and astronomical surveys would have to close down, including CERN and LIGO. 

\cite{Watters2026} argument conflates two fundamentally different methodological frameworks: object-level validation, as used in the construction of machine-learning training datasets, and ensemble-level statistical inference. In the latter framework, conclusions do not depend on the physical verification of individual detections but instead on whether any contamination present is systematically correlated with the variable being tested. 

This allows, for example, large population studies to be conducted even in the presence of imperfect or noisy data. In statistical inference, the sample is chosen to balance false positives and false negatives, so as to avoid both the inclusion of artefacts incorrectly classified as real and the erroneous removal of genuine signals. This is particularly important because many filtering choices can skew the sample.

The aggressively filtered sample of 5{,}399 transients presented in \cite{Solano2022} was constructed for a \textit{different scientific purpose}: the search for true vanishing stars, defined as objects detected in the 1950s that cannot subsequently be recovered in any catalogue, see \cite{Villarroel2020}. To minimise false positives in that context, a strict requirement was imposed that no cross-match with any other astronomical catalogue could exist across the electromagnetic spectrum, after which scanning artefacts were removed. 

\begin{figure*}[t]
\centering

\begin{subfigure}{0.48\textwidth}
    \centering
    \includegraphics[width=\linewidth]{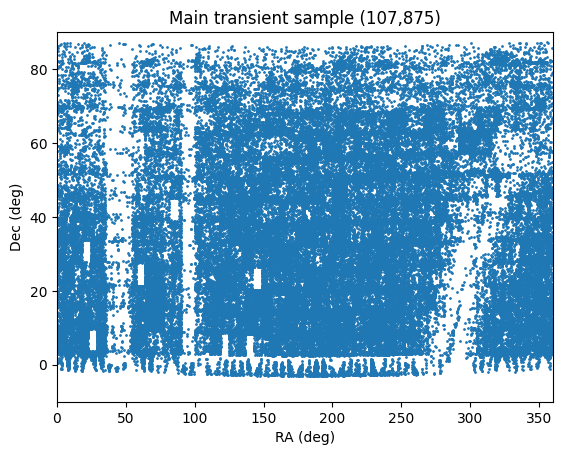}
\end{subfigure}
\hfill
\begin{subfigure}{0.48\textwidth}
    \centering
    \includegraphics[width=\linewidth]{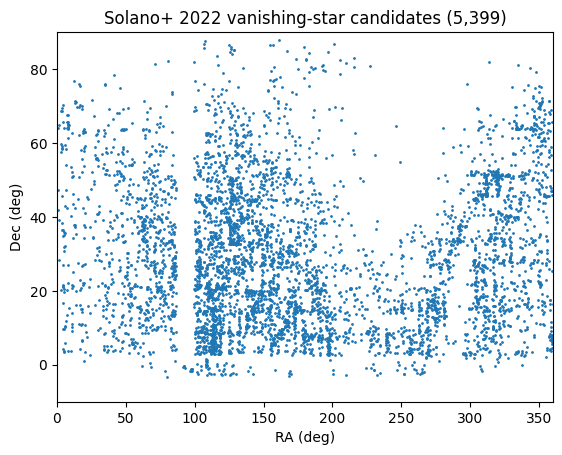}
\end{subfigure}

\medskip

\begin{subfigure}{0.48\textwidth}
    \centering
    \includegraphics[width=\linewidth]{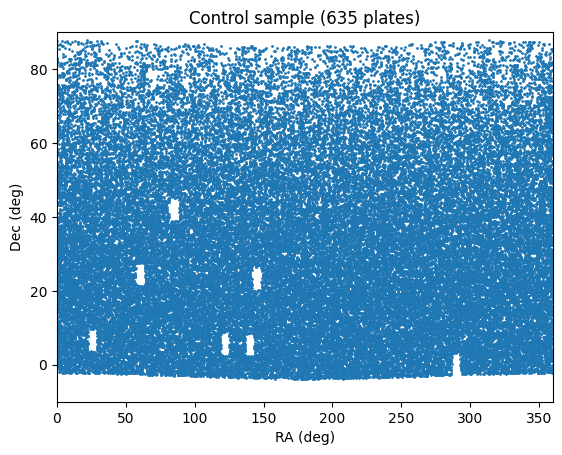}
\end{subfigure}
\hfill
\begin{subfigure}{0.48\textwidth}
    \centering
    \includegraphics[width=\linewidth]{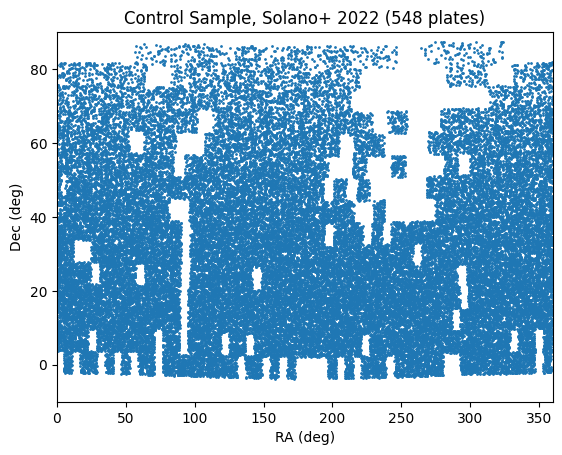}
\end{subfigure}

\caption{\textbf{Sky coverage.} Four different scenarios are shown: (upper left) the main transient sample of 107,875 objects; (upper right) the aggressively filtered sample used in Watters et al. (2026), also cross-matched against every existing astronomical catalogues across the electromagnetic spectrum; (lower left) the sky coverage of the plates included in the main transient sample; and (lower right) the sky coverage of the plates included in the aggressively filtered sample. The aggressively filtered control misses a large central region of the sky, resulting in an inherent excess of transients relative to the more homogeneous control sample of the main dataset. We further examine whether the stripes in the main transient sample influence the result by restricting the analysis to $100^\circ < \mathrm{R.A.} < 250^\circ$. Instead of explaining the deficit, excluding these regions leads to a stronger signal.}
\label{fig:four_panels}
\end{figure*}

Given that the catalogues used for cross-matching have different sky coverages, this introduces heterogeneity into the aggressively filtered sample. We illustrate this heterogeneity in Figure \ref{fig:four_panels}. A substantial spatial region is entirely missing from the aggressively filtered dataset, which becomes particularly evident when examining the corresponding control sample constructed from its sky coverage. The control sample does not show the position of each transient, but rather which plates are part of the sky coverage. The aggressively filtered sample is therefore strongly spatially biased as a result of the additional cross-matching against multiple astronomical catalogues. This contrasts with the main transient sample, where variations in the transient distribution do not simply reflect differences in sky coverage, as demonstrated by the homogeneous control sample.
By comparing the shadow fractions of the two control samples—thereby isolating the effect of sky coverage alone—we find that the aggressively filtered sky coverage produces an $\sim 17\%$ excess of transients in the Earth’s shadow. Given the strong spatial bias in the sky coverage, the smaller sample is not well suited for testing the Earth-shadow effect.

The 5{,}399  catalogue was therefore not created for completeness but for the search for a candidate vanishing star. Upon further investigation, however, none of these vanishing star-candidates was shown to be a genuine vanishing star; instead, the remaining candidates were interpreted as elusive transient events. Anyone using the 5{,}399 transients is also recommended to add similar additional filtering steps, corresponding to what was done for the 298,165 to $\sim$ 107,000 objects, to remove potential duplicates and sources that escaped the Gaia and PanSTARRS filtering.

When \cite{Watters2026} select this aggressively filtered sample, they do not account for the fact that the transients surviving such filtering are those that remained after a specific search for vanishing stars. As such, this selection misses a much larger population of short-lived transients present in the Palomar plates. This is also why we initiated our analysis using a much larger set of transients.

When the aim is to search directly for short-lived transients present in a single plate exposure, such aggressive filtering is not well motivated. First, artificial or non-standard objects need not be strictly absent at all other epochs. Depending on geometry, illumination conditions, and surface properties, an object may appear only as a brief optical glint once while remaining faint or undetectable in other bands or at other times. For example, a reflective surface may produce a transient optical flash under specific solar alignment, while otherwise appearing as a marginal or infrared-dominated source. Excluding every transient that may recur at nearly the same position, or that exhibits a faint counterpart in any existing astronomical survey across the electromagnetic spectrum, implicitly imposes assumptions about the physical nature and persistence of the source that are not required by the search hypothesis. For example, filtering out a visual transient because catalogs indicate a non-visible, faint infrared object in the same region (which could be coincidental) is not warranted when one is seeking to identify visual transients. Such criteria therefore risk discarding genuine signals.

In this context, it is methodologically preferable to err on the side of false positives and survey completeness rather than on the side of false negatives.

The sample of $\sim$107,000 transients \footnote{Courtesy of Enrique Solano.} used in \citep{VillarroelPASP,Bruehl2025} is an improved version of the 298,165 objects used in \cite{Solano2022}. The 298,165 objects are an intermediate, unpublished sample used in the workflow, based on morphometric and kinematic parameters. It contained the southern hemisphere (including a complementary catalogue for the southern hemisphere), many duplicates and some Gaia and PanSTARRS sources that despite previous filtering remained in the catalogue. For statistical inference, the catalog was revisited during early 2025. For the 298,165 objects it was again required no counterpart within 5{''} in Gaia and/or Pan-STARRS, removed all duplicates and all objects from the southern hemisphere. The \cite{VillarroelPASP} article contains an inaccurate statement indicating that NeoWISE crossmatches were removed; NeoWISE crossmatches were not part of the filtering procedure. Thus, the ~107,000-object sample is far more appropriate for statistical inference than the 5,399-object subset used by Watters et al., both because of its substantially larger size, removal of certain false positives and because it avoids cross-matching against a zoo of catalogues across the electromagnetic spectrum with differing and uneven sky coverages.

As we commented upon in the same publication, the sample has not been visually inspected.
As expected, such a sample might include plate defects, scanning artefacts, and variable stars, which together constitute statistical noise. However, this approach allows the use of the largest possible dataset, enabling searches for alignments and other subtle statistical effects. In \cite{VillarroelPASP} Section 8, we estimated that approximately one third of the more than 107,000 transients are solar reflections, while up to two thirds may be false positives, including plate defects. What matters is not whether our sample contains noise, but whether any contamination is systematically correlated with the tested variable. The limitations of our sample are discussed in Section~2 of \cite{VillarroelPASP}.

To assess the relative data quality of the samples used in previous work, we performed a cross-match between the 5,399-object catalogue of \cite{Solano2022}, later adopted by \cite{Watters2026}, and the larger VASCO sample of 107,875 transients. The latter represents a post-vetted version of the original $\sim$298,000 detections, in which a substantial fraction of artefacts had already been removed. Using a positional tolerance of $<1$ arcsecond, we find that only 1,842 objects from the 5,399 catalogue are present in the vetted 107,875 sample, implying that a significant fraction of the smaller catalogue consists of spurious detections that do not survive independent validation.

In fact, it can readily be shown that the aggressive filtering to obtain 5,399 objects does not necessarily dramatically remove false positives from the dataset. As an illustrative comparison, 10 objects from the 107,000-transient sample and 10 objects from the aggressively filtered sample of 5,399 objects presented in \cite{Solano2022} were randomly selected and inspected visually. In such a comparison, 4 of the 10 objects drawn from the 107,000 sample look like transients, while 5 of the 10 objects from the aggressively filtered sample do so. While this small illustrative exercise does not constitute a formal statistical test, it does not indicate a dramatic improvement in purity commensurate with the approximately twenty-fold reduction in sample size. Given that the latter sample is reduced in size by approximately a factor of 20, and that the aggressive filtering does not appear to remove false positives (but instead addresses a different physical question), there is no clear advantage in adopting it for the present analysis.

We then compared the machine-learning classification probabilities assigned within the 107,875-sample for these matched objects against the full population, across a range of probability thresholds ($p > 0.2$–$0.9$), see \textcolor{blue}{Bruehl et al. (2026)}. Only 1,872 of the 5,399 transients in the aggressively filtered sample overlap with the 107,875-object catalogue. These were used to estimate the false positive rate. The results, summarised in Table \ref{tab:comparison}, show that the matched subset consistently exhibits a lower fraction of high-probability detections than the full sample. This demonstrates that the catalogue used by \cite{Watters2026} is not a cleaner subset of the data; rather, despite its smaller size, it contains a higher relative fraction of low-confidence and likely spurious events.

\begin{table}
\centering
\caption{\textbf{Comparison of probability distributions.} The table lists the fraction of objects in each sample that pass a given probability threshold for being visually classified as transients, based on the machine learning classifier of \textcolor{blue}{Bruehl et al. (2026)}. We show results for the full VASCO sample ($N = 107{,}875$) and the matched subset of the aggressively filtered sample ($N = 1{,}842$).}
\label{tab:comparison}
\begin{tabular}{ccc}
\hline
Threshold ($p$) & Full sample fraction & Matched subset fraction \\
\hline
0.2 & 0.621 & 0.502 \\
0.3 & 0.504 & 0.400 \\
0.4 & 0.409 & 0.325 \\
0.5 & 0.324 & 0.257 \\
0.6 & 0.249 & 0.209 \\
0.7 & 0.171 & 0.134 \\
0.8 & 0.082 & 0.055 \\
0.9 & 0.013 & 0.012 \\
\hline
\end{tabular}
\end{table}

In our case, one of our tested variables is Earth-shadow geometry. We observe a strong, statistically significant deficit of transients within the Earth’s shadow at the $\sim 7.6\sigma$ level. When using the full set of $\sim$107{,}000 transients, the Earth-shadow deficit corresponds to a significance of $22\sigma$ when compared to the theoretical hemispherical coverage, and approximately $7.6\sigma$ when evaluated using a sky-coverage-based approach against a control sample. When restricting the analysis to only transients within 2 degrees from the plate center (to avoid plate edge artifacts), the sky-coverage-based deficit remains at $\sim$30\%, but with a statistical significance of $2.6\sigma$ solely due to the effects of reduced statistical power.

Random noise, plate defects or scanning artifacts, or spurious detections may add background, but they cannot generate a global, directional deficit aligned with the Earth’s shadow. Such defects are plate-fixed and time-independent, whereas the Earth’s shadow is time-dependent and sky-projected. Stars, on the other hand, reside far, far outside the shadow cone. For plate or scanning defects to generate a deficit in the Earth’s shadow, they would need to develop consciousness and an unexpected fondness for orbital mechanics. Similarly, for stars to systematically avoid the Earth’s shadow, one would have to adopt exotic new assumptions about physics. And as previously mentioned, we see a strong and significant deficit of transients in the Earth's Shadow, both when comparing to the theoretical hemisphere coverage and the actual sky coverage.

\cite{Watters2026} report no deficit when analysing a reduced subset of 4{,}866 transients drawn from the original 5{,}399-object sample of \cite{Solano2022}. This issue easily boils down to the problem of not having enough statistical power.

For a fixed underlying effect size, the expected significance scales approximately with the square root of the sample size. Starting from a $7.6\,\sigma$ deficit in the full sky cover analysis of 107{,}875 transients, a reduction to 4{,}866 objects implies an expected significance of order\[
7.6 \times \sqrt{\frac{4866}{107875}} \approx 1.6\,\sigma,
\] even before accounting for additional variance introduced by plate-by-plate solid-angle modelling. 

The analysis presented in \cite{Watters2026} further contains no error bars, uncertainty estimates, or error propagation. In addition, their solid-angle analysis includes no modelling of how methodological choices—such as cell size—affect the resulting uncertainties.

It is therefore unsurprising that no statistically significant deficit is recovered when the analysis is performed on a sample of only 4{,}866 transients and without any understanding of the accompanying uncertainties. In such a regime, non-detections or apparent excesses are not informative about the presence or absence of the effect, but are instead the statistically expected outcome of an underpowered test.

Simply speaking, the sample used in \cite{Watters2026} is not appropriate for testing our finding of an Earth-shadow deficit reported in \cite{VillarroelPASP}, as its sample size is too small for that purpose.

The next fundamental issue in \cite{Watters2026} is the use of the \cite{Solano2022} sample as these data lack fundamental information on observation time and date\footnote{\url{http://svocats.cab.inta-csic.es/vanish-possi/index.php?action=search}}. Without this information, the position of the Earth’s shadow cannot be correctly computed, nor can any temporal analysis be considered valid. Although \cite{Watters2026} acknowledge this limitation, they attempt to mitigate it by assigning to each transient the closest plate in horizontal separation. In addition, all transients associated with overlapping plates are removed, since the same region of sky may have been covered multiple times. This procedure is highly sensitive to any uncertainties, which have not been quantified, and any guessing on the wrong ``plate centre'' during horizontal separation calculations will lead to misassignment. It is not demonstrated that the assigned observation dates and times correspond to the correct transient events, and how big fraction of the transients have been assigned to the wrong plate and observation time.

\begin{figure*}
   \includegraphics[scale=0.3]{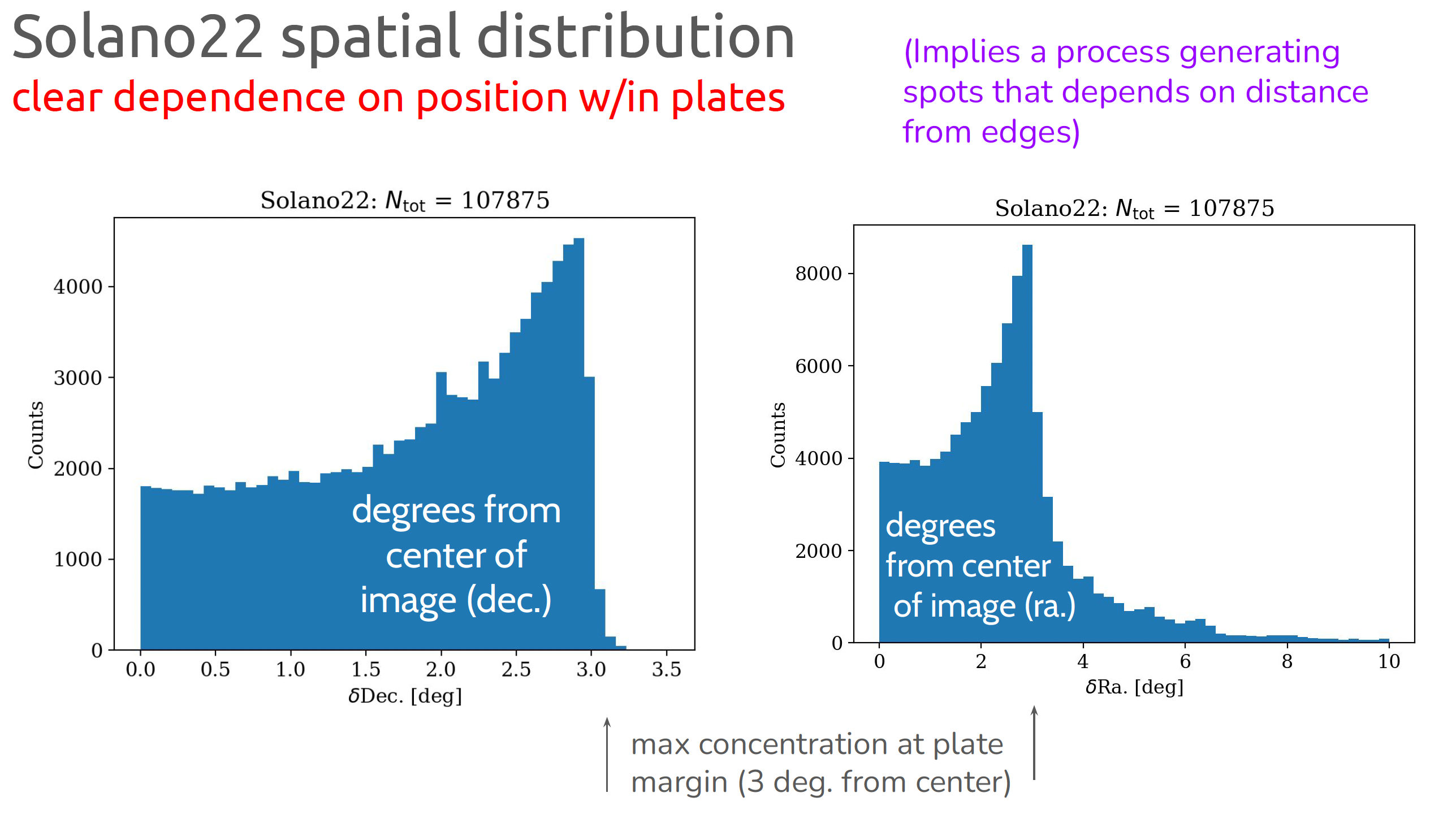}
  \caption{\label{histogram} {\bf Illustrative histogram of the 107,000 transients.} In the last communication between W.~Watters and B.~Villarroel in July 2025, Watters shared this central figure with above text to express his concern regarding an apparent increase in the number of transients towards the plate edge (see the \textit{Acknowledgements} in Watters et al.~2026). The photographic plates are $6\times6$ degrees in size, whereas the x-axis of the histogram extends to 10 degrees from the plate centre, and outside the plate edge. This suggests missing a cos(Dec)-correction when calculating the horizontal separation (difference in R.A.). Further, the histogram does not take into account plate overlap regions. The figure serves to illustrate how sensitive the time reconstruction in Watters et al. (2026) is to the adopted horizontal separation estimates and definitions. We thank W.~Watters for sharing this figure.}
   \end{figure*}

\begin{figure*}
   \includegraphics[scale=0.5]{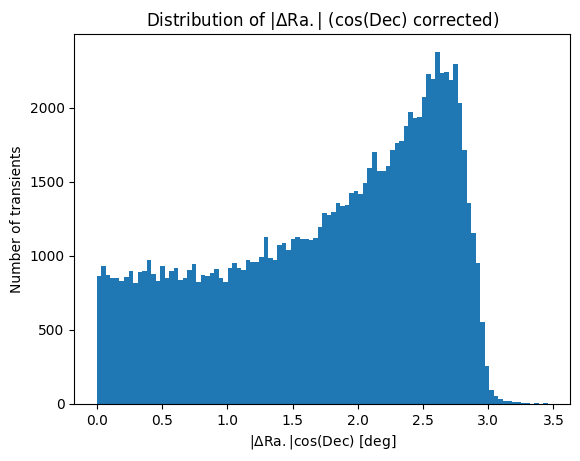}
   \includegraphics[scale=0.5]{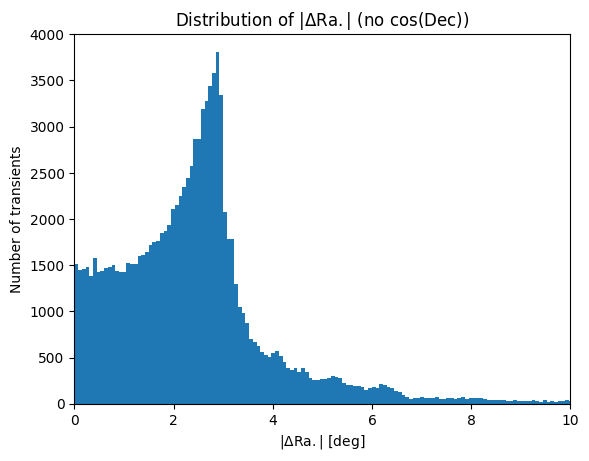}
  \caption{\label{histogram2} {\bf Histogram with and without cos(Dec)-correction.} The left figure (correct) shows a histogram with cos(Dec) correction when calculating the horizontal separations. The right figure shows a histogram without the cos(Dec) correction when calculating the horizontal separations from the plate center. The right distribution looks similar to Watters figure, albeit with narrower bins.}
   \end{figure*}

This concern is easiest illustrated by a brief examination of a histogram presented by W.~Watters\footnote{Based on the last email correspondence between the two first authors of the respective papers, Villarroel, B.\ and Watters, W., July 2025.}, showing the number of transient candidates as a function of horizontal separation (in degrees) from the plate centre. The histogram is constructed using the 107,000 transients with well-determined observation dates and times from \cite{VillarroelPASP} (see Fig.~\ref{histogram}). We note that the photographic plates cover only a box of $6\times6$ degrees, whereas the horizontal axis of the histogram extends to offsets of up to 10 degrees from the plate centre. This suggests the horizontal separation in this plot is missing the cos(Dec)-correction factor, see (see Fig.~\ref{histogram2}).
\footnote{In correspondence in February 2026, W.W. stated that no such error had occurred and requested removal of the figure. However, the figure and its analysis are still referenced in the Acknowledgements of Watters et al. (2026) when mentioning “early concerns” and “distributions of SPFs”. We note that his concerns (including others) were directly addressed in Sections 2 and 8 of Villarroel et al. (2026), forming the basis for both the $7.6\sigma$ result and the test restricted to transients near the centre of the plate to minimise plate defects. The shadow deficit remained significant.}

This raises questions about how the horizontal separation in time reconstruction is actually defined in \cite{Watters2026}, including whether a cos(dec) factor enters the calculations. Horizontal separations are used in Section 4.1 of  \cite{Watters2026} to infer the times and dates. But if the horizontal separations are not reliably computed, the association between a transient and the correct photographic plate—and hence the correct observation time—breaks down at higher declinations. Since nearly half the plates lie at Dec. $\ge 30^\circ$, this would introduce a substantial multiplicative distortion across a large fraction of the dataset and influence every time and date estimate. All inferences, such as Earth Shadow fractions and correlations with nuclear weapon tests and UFOs -- would be incorrect. While the detective-style approach presented in \cite{Watters2026} is certainly meticulous, without a quantitative assessment of uncertainties and of the bias introduced by the method, the reduced sample used in \cite{Watters2026} cannot reasonably be regarded as ``validated'' for a test that is extremely sensitive to precise time and date assignments. In contrast, the sample of 107,000 transients includes explicit observation times and dates for each event.

Taking into account plate overlap among POSS-I plates and the survey geometry through a Monte Carlo simulation, we note that the excess relative to an isotropic expectation becomes significant only at radial distances $\gtrsim 2.5^\circ$ from the plate centre. The test performed in \cite{VillarroelPASP}, where the shadow fraction was evaluated within $2^\circ$, is therefore not affected by this edge excess and remains valid.

The next issue concerns the temporal correlations with nuclear weapon tests and UAP in \cite{Bruehl2025}. The critique by \cite{Watters2026} \ (Section~6) highlights the relatively small number of nights with Palomar observations, noting that the published article conservatively assumed zero transients on nights for which no transient was reported. The authors conclude that the apparent associations between transients and nuclear testing arise solely from coincidental overlap between nights of Palomar observation and nuclear testing schedules.

To address this concern, we downloaded the POSS-I observation dates and merged these with our analysed nuclear--transient dataset (observation dates from: \url{https://vizier.cds.unistra.fr/viz-bin/VizieR} (VI/25/nposs)). We then analysed the dataset restricted to the 370 nights with confirmed POSS-I observations. While at least one transient was observed on 72.8\% of dates outside a nuclear testing window ($\pm$1~day), transients were observed on 84.2\% of dates within a nuclear testing window. This $\approx$11\% absolute difference is substantially larger than the 4.8\% difference reported in the original paper. The exact one-tailed $p$-value (testing a directional hypothesis) for the dichotomous association between transients and nuclear testing is $p = 0.046$. A non-parametric test likewise showed a significantly greater number of transients (Mann--Whitney $U$, $p = 0.046$) within a nuclear testing window compared to outside such a window. A Poisson generalised linear model analysis further revealed significantly more transients on POSS-I observing nights within, as opposed to outside, a nuclear testing window ($p < 0.001$). The reported correlation between the number of transients and the number of UAP sightings per night was essentially unchanged from our paper ($\rho_{\mathrm{Spearman}} = 0.13$, $p = 0.011$). Finally, the analysis demonstrating that being within a nuclear testing window and having at least one UAP sighted on a given night produced additive effects on the number of observed transients remained highly significant ($p < 0.001$), with the overall pattern unchanged. Thus, even when using the restrictive dataset suggested by \cite{Watters2026}, the results reported in the \textit{Scientific Reports} paper remain unchanged. We further note that while coincidental timing between POSS-I observation nights and nuclear testing schedules might hypothetically contribute to associations between dichotomous occurrence of transients (presence/absence) and nuclear testing, it cannot account for findings that the actual number of transients is significantly higher when within a nuclear window. This issue is not addressed by the \cite{Watters2026} critique.

The results in \cite{Bruehl2025} results have further been replicated by two professional data analysts. While not peer-reviewed, the results are consistent with our findings. See, for example, an independent analysis available online conducted by Janne Sinkkonen which controls for temporal clustering of observations \footnote{https://github.com/euxoa/plates}.

In their MAPS-based analysis, \cite{Watters2026} compare the spatial properties of the VASCO transients to those of objects drawn from the MAPS catalogue, which consists of sources detected consistently across multiple photographic plates and bandpasses. By construction, MAPS therefore represents a population of longlived astronomical transients and excludes short-lived phenomena that appear in only a single exposure (seconds timescales). Deviations between the spatial distributions of MAPS objects and the VASCO transients are interpreted as evidence that the latter are dominated by plate or scanning artefacts. However, this comparison is not informative for the hypothesis under consideration. The VASCO transients consist of short-duration optical transients, including phenomena that are not expected to persist across multiple exposures separated by hours or days. Using a catalogue that systematically removes such events as a reference population therefore precludes the very class of signals being investigated. Similar category errors arise in discussions of early searches for optical counterparts to Gamma-Ray Bursts, an exercise we leave to the reader.

In summary, the 30-page critique by \cite{Watters2026} offers a detailed and systematic examination of the analyses presented in \cite{VillarroelPASP} and \cite{Bruehl2025}. It makes a case for dismissing any sample with a large fraction of noise, but just as the Higgs boson could only be found by embracing the statistical power of a vast and noisy ensemble, we must not allow a hyper-fixation on individual plate defects to obscure the clear, large-scale signals that remain imprinted on the historical sky.

While \cite{Watters2026} level of scrutiny can be valuable and motivates us for improving our methodology, the work by \cite{Watters2026} is undermined by a lack of careful alignment between the hypotheses being tested and the methodological frameworks and samples employed to test them. When combined with an inadequate treatment of uncertainties and statistical power, no accurate information on times and dates of the events, several of the resulting conclusions stem from misunderstandings. Taken together, the critique serves as a reminder that the present analysis establishes a robust starting point, not an endpoint. For now, the transients remain on the plate.



\section{Acknowledgments}
B. Villarroel is grateful to Piotr Tchaikovsky for inspiration.


\bibliographystyle{mnras}
\bibliography{ss-seti_bv.bib}




\bsp	
\label{lastpage}
\end{document}